\documentclass[aps,pra,twocolumn,superscriptaddress,showpacs,10pt]{revtex4-1}
\usepackage{times,amsmath,amssymb,amsfonts,graphicx}
\newcommand{\ket}[1]{\vert #1 \rangle} %
\newcommand{\bra}[1]{\langle #1 \vert} %

\newcommand{\ketbra}[2]{\vert #1 \rangle \! \langle #2 \vert}

\def\Tr{\hbox{Tr}}
\begin{document}
\title{Experimental estimation of quantum discord for polarization qubit 
and the use of fidelity to assess quantum correlations}
\author{Claudia Benedetti}
\affiliation{Dipartimento di Fisica, Universit\`a degli 
Studi di Milano, I-20133 Milano, Italy} %
\author{Alexander P. Shurupov}
\affiliation{INRIM, I-10135, Torino, Italy}
\author{Matteo G. A. Paris}
\affiliation{Dipartimento di Fisica, Universit\`a 
degli Studi di Milano, I-20133 Milano, Italy} %
\author{Giorgio Brida}
\affiliation{INRIM, I-10135, Torino, Italy}
\author{Marco Genovese}
\affiliation{INRIM, I-10135, Torino, Italy}
\date{\today}
\begin{abstract}
We address the experimental determination of entropic quantum discord for systems
made of a pair of polarization qubits.  We compare results from full and
partial tomography and found that the two determinations are
statistically compatible, with partial tomography leading to a smaller
value of discord for depolarized states.  Despite the fact that our
states are well described, in terms of fidelity, by families of
depolarized or phase-damped states, their entropic discord may be largely
different from that predicted for these classes of states, such that no
reliable estimation procedure beyond tomography may be effectively
implemented.  Our results, together with the lack of an analytic
formula for the entropic discord of a generic two-qubit state, demonstrate that
the estimation of quantum discord is an intrinsically noisy procedure.
Besides, we question the use of fidelity as a figure of merit to assess
quantum correlations.
\end{abstract} 
 \maketitle
\section{Introduction} 
Quantum correlations are central resources for quantum technology.  In
the recent years, it has become clear that besides entanglement \cite{G}
novel concepts may be introduced to capture  more specific aspects, such
as quantum steering or quantum discord \cite{rev1,rev2}.  Quantum
discord  has recently attracted considerable attention
\cite{dis,vogl,rahimi,galve,lars,bla12,haikka,vas10,Yu2007,dakic,gio10,fer12,
chuan,perinotti,allegra,bellomo,mazzola,chiuri,daoud,usha,Xu10,add1,add2,add3}, 
as it
captures and quantifies the fact that quantum information in a bipartite
system cannot be accessed locally without causing a disturbance, at
variance with classical probability distributions.  Yet, the relevance
of quantum discord as a resource is a highly
debated topic \cite{Rod08,Bro13}, and a definitive answer  may only come
from experiments involving carefully prepared quantum states. This poses
the problem of a precise characterization of quantum discord and of the
design of optimized detection schemes \cite{bla12,ade13}.
\par
For a given quantum state $\rho$ of a bipartite system $AB$, the total amount of
correlations is defined by the quantum mutual information%
\begin{equation}\label{eq:QMI}
I(\rho) = S(\rho_A) + S(\rho_B) - S(\rho),
\end{equation}
where $S(\rho) = -\Tr[\rho \log \rho]$ denotes von Neumann entropy and
base two logarithm is employed. An alternative version of the quantum
mutual information, that quantifies the classical
correlations, is defined as
\begin{equation}\label{eq:CCor}
J_A = S(\rho_B) - \min \sum_k p_k S(\rho_{B \vert k}).
\end{equation}
where $\rho_{B \vert k}=\text{Tr}_A[\Pi_k\rho\Pi_k]/
\text{Tr}[\Pi_k\rho\Pi_k]]$ is the conditional state of system 
$B$ after obtaining outcome $k$ on $A$, $\left\{\Pi_k\right\}$ are 
a projective measurements on $A$, and $p_j=\text{Tr}[\Pi_k\rho\Pi_k]]$
is the probability of obtaining the outcome $k$.
While these two definitions are equivalent in classical information, 
the difference between them in quantum case defines entropic 
quantum discord
\begin{equation}\label{eq:Disc}
D_A(\rho) = \mathcal{I}(\rho) - \mathcal{J}_A(\rho).
\end{equation}
An analogue quantity may be defined upon performing measurements on the
system $B$. Notice also that different quantities, based on distances 
rather than entropy, have been proposed to measure
quantum correlations \cite{dak10,pia12}, and recently measured experimentally without 
the need of tomographic reconstruction of the density matrix \cite{ade13}.
\par
In this paper we address estimation of discord for different families of
mixed states obtained from initially pure maximally entangled states.
Our purpose is that of understanding what kind of measurements are
really needed for a precise determination of this quantity. In
particular, our setup is designed such to generate depolarized and/or
phase-damped version of polarization Bell states 
$\ket{\Phi^{+}}$ and $\ket{\Psi^{+}}$. The first class of states is 
that of Werner (depolarized) states:
\begin{equation}
\label{eq:MStateW}
\rho_W(p) = p\ketbra{\Psi}{\Psi} + (1-p) 
\frac{\mathbb I}{2}\otimes\frac{\mathbb I}{2}
\end{equation}
where $\ket{\Psi}=\ket{\Phi^{+}},\ket{\Psi^{+}}$, $\mathbb I$ is 
the identity matrix, and $p$ is a mixing parameter 
related to the purity of $\rho_W$ by 
the formula $\mu= \Tr\rho_W^2 = (1 + 3 p^2)/4$. The second class 
of states we are going to investigate is given by phase-damped 
(decohered) version of Bell states, i.e. 
\begin{equation}\label{eq:MStateDphi}
\rho_{D}(p) = p\ketbra{\Psi}{\Psi} + (1-p)\sum_{kj} P_{kj}
\ketbra{\Psi}{\Psi}P_{kj}\,,
\end{equation}
where $\ket{\Psi}=\ket{\Phi^{+}},\ket{\Psi^{+}}$, 
$P_{kj}=|k\rangle\langle k|\otimes |j\rangle\langle j|$ are 
the diagonal projectors over the standard basis, and the parameter $p$ is related 
to the purity by the relation $\mu= (1 + p^2)/2$.
\par
Both the families (\ref{eq:MStateW}) and (\ref{eq:MStateDphi}) 
belong to the class of $X$-states (named after the shape of the
nonzero portion of the density matrix). In addition, they may 
be written as 
\begin{equation}\label{eq:Luo} \rho = \frac{1}{4} \left( \mathbb I +
\sum_{j=z,y,z} c_j \sigma_j \otimes \sigma_j \right) \end{equation}
where  $\sigma_j$ are Pauli
operators. Werner states
(\ref{eq:MStateW}) are obtained for
$c_1=-c_2=c_3=p$, while decohered states
(\ref{eq:MStateDphi}) correspond to $c_1=-c_2=p$,
$c_3=1$. 
For bipartite states of the form (\ref{eq:Luo}) a general analytic formula 
for the discord has been obtained \cite{Luo}, leading to
\begin{align}
\label{dw}
D(\rho_W)=&\frac14\left[3(1-p) \log (1-p) + (1+3p) \log(1+3p)\right]
\notag \\  &- \frac12 \left[(1-p) \log (1-p)+(1+p) \log (1+p)\right] \\ 
\label{dd}
D(\rho_D)= & \frac12\left[(1-p) \log (1-p) + (1+p) \log(1+p)\right]
\end{align}
where we omit to denote the measured party since 
both families are made of symmetric states.
For decohered states the optimal measurement to access the classical 
correlations is given by polarization measurement along the z-axis,
whereas the fully symmetric structure of Werner states makes all the
Pauli operators optimal.
\par
Discord is a nonlinear functional of the density matrix and cannot, even
in principle, correspond to an observable in strict sense. Its
determination thus unavoidably involves an estimation procedure from 
a suitable set of feasible measurements \cite{LQE}.
In the following 
we present our experimental results about estimation of discord via tomographic
reconstruction and investigate the possibility of its determination by a
restricted set of measurements. We found that despite the fact that our
states are well described, in terms of fidelity, by families of
depolarized or phase-damped states, their discord may be largely
different from that predicted for these classes of states, such that no
reliable estimation procedure beyond tomography may be effectively
implemented.  Our results, together with the lack of an analytic
formula for the discord of a generic two-qubit state \cite{gir11}, 
demonstrate that the estimation
of quantum discord is an intrinsically noisy procedure. 
\par
In the next Section we describe our experimental setup as well as our
tomographic reconstruction. The estimation of discord from experimental
data is described in Section \ref{s:expD} whereas the discussion of
results is reported in Section \ref{s:res}.
 Section \ref{s:out} closes the paper with some concluding
remarks.
\section{Experimental setup and reconstruction of the density matrix}
\begin{figure}[h!]
\includegraphics[width= 0.95\columnwidth]{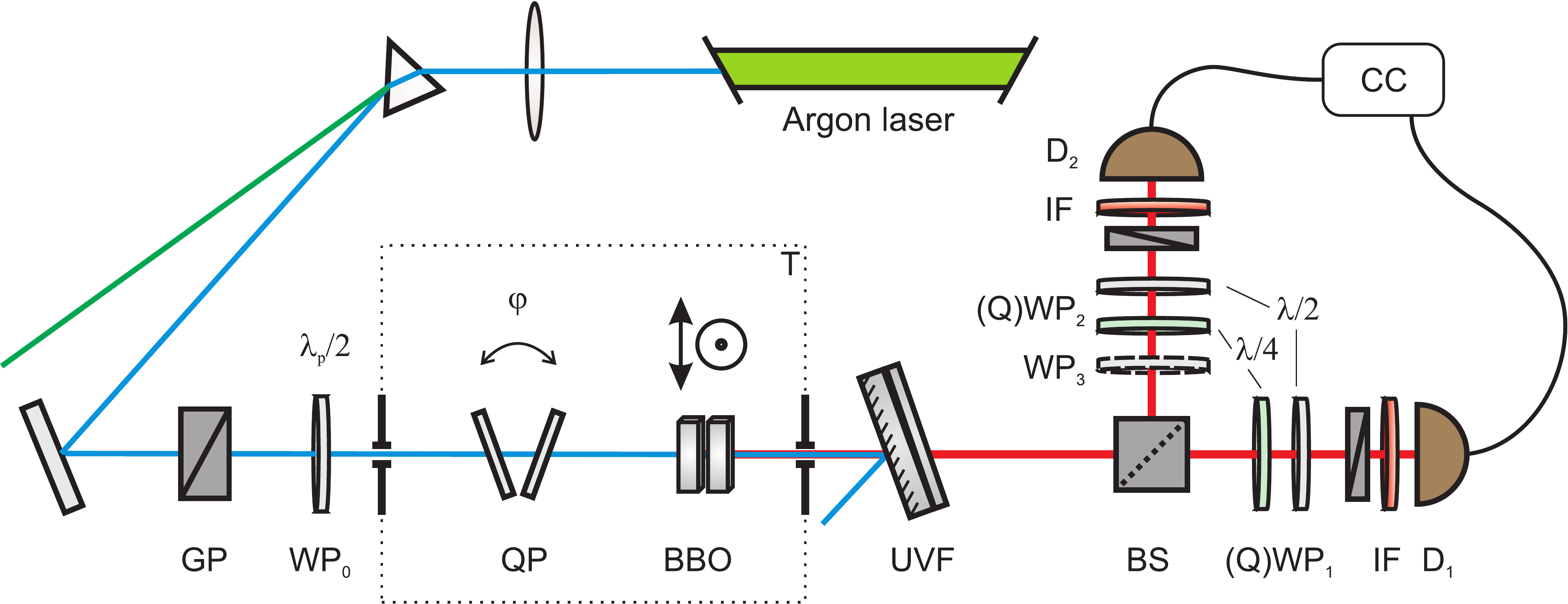}
\caption{(Color online) Experimental setup to generate polarization
photon pairs with variable quantum correlations and to perform 
tomographic reconstruction of the generated state.} \label{f:setup}
\end{figure}

In our setup polarization two-qubit states are generated using 
parametric down-conversion in a two-crystal geometry. In an optical 
crystal with type-I non-linearity, one photon of the pump beam decays into 
a pair of photons having same linear polarization. In our experiment 
(see Fig.~\ref{f:setup}) we have used Argon laser beam at 351~nm to 
pump two type-I beta barium borate (BBO) crystals fixed to have their optical
axes in perpendicular planes. 
\par
A Glan-Thompson prism (GP) is used to project initial laser
beam polarization to horizontal plane. Half-wave plate (WP$_0$) 
is then used to rotate at $45^\circ$ the polarization of the pump 
beam. First (second) BBO crystal has its optical axis in
horizontal (vertical) plane and produce a pair of photons 
when pumped by a horizontally (vertically) polarized beam.
Due to the coherent superposition of the two emissions the setup is 
suitable to create polarization entangled states of the form:
\begin{equation}
\ket{\Psi_{\theta,\varphi}} = \cos\theta\ket{HH}+ e^{i\varphi}
\sin\theta\ket{VV} \label{psi:th:ph}
\end{equation}
The quartz plates (QP) can be tilted to tune the phase between the two
components. In the experiments reported in this paper we have fixed it
to zero. We also set the superposition angle to $\theta=\pi/4$ in order to
generate maximally entangled states. Non-linear crystals and quartz
plates are placed in to a temperature-stabilized closed box for
achieving stable phase-matching conditions all the time. 
\par 
The portion of our setup devoted to the characterization of the
two-qubit states starts with a beam splitter (BS), which 
is used to split the initial (collinear) biphoton field 
into two spatial modes. In each output arm a quarter--wave (QWP$_i$) 
and a half--wave (WP$_i$) plates are placed, and followed by 
a linear polarizer and and interference filter (IF) with
central wavelength at 702~nm (FWHM 10~nm). Avalanche photodiode 
detectors (D$_i$) are placed and the end and connected to a 
coincidence count scheme (CC).
\par
In order to prepare basis states with both photons having vertical
(horizontal) polarization we have rotated the half--wave plate WP$_0$ to
feed only first (or second) crystal. To prepare basis states with
orthogonal polarization of photons an additional half--wave
plate(WP$_3$) may be introduced  in the reflected arm. In the same way we
are able to transform the initial
$\ket{\Phi^+}=\frac{1}{\sqrt{2}}(\ket{HH}+\ket{VV})$ to
$\ket{\Psi^+}=\frac{1}{\sqrt{2}}(\ket{HV}+\ket{VH})$ state.
In addition, upon mixing maximally entangled and basis states
obtained with different measurement times, we have analyzed Werner and phase-damped 
states with purity $\mu$ equal to $0.25$, $0.5$, $0.67$, $0.83$ and $1$. 
\begin{figure}[h!]
\centering
\includegraphics[width=0.45\columnwidth]{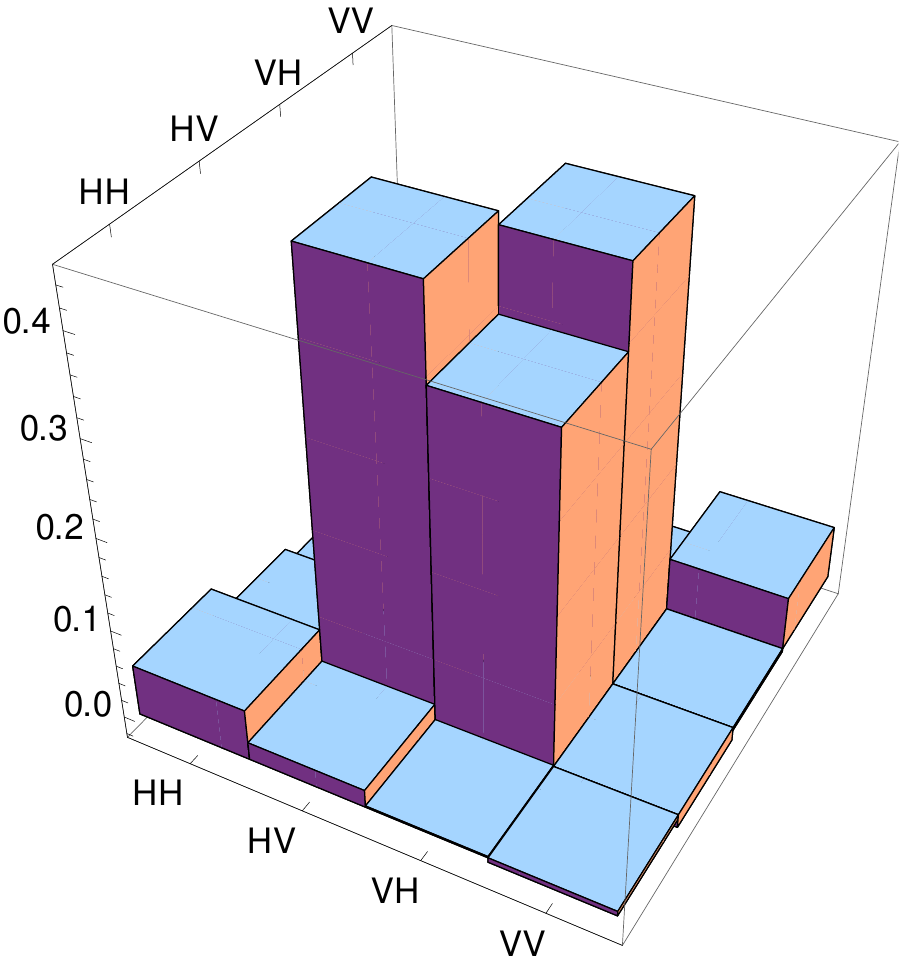}
\includegraphics[width=0.45\columnwidth]{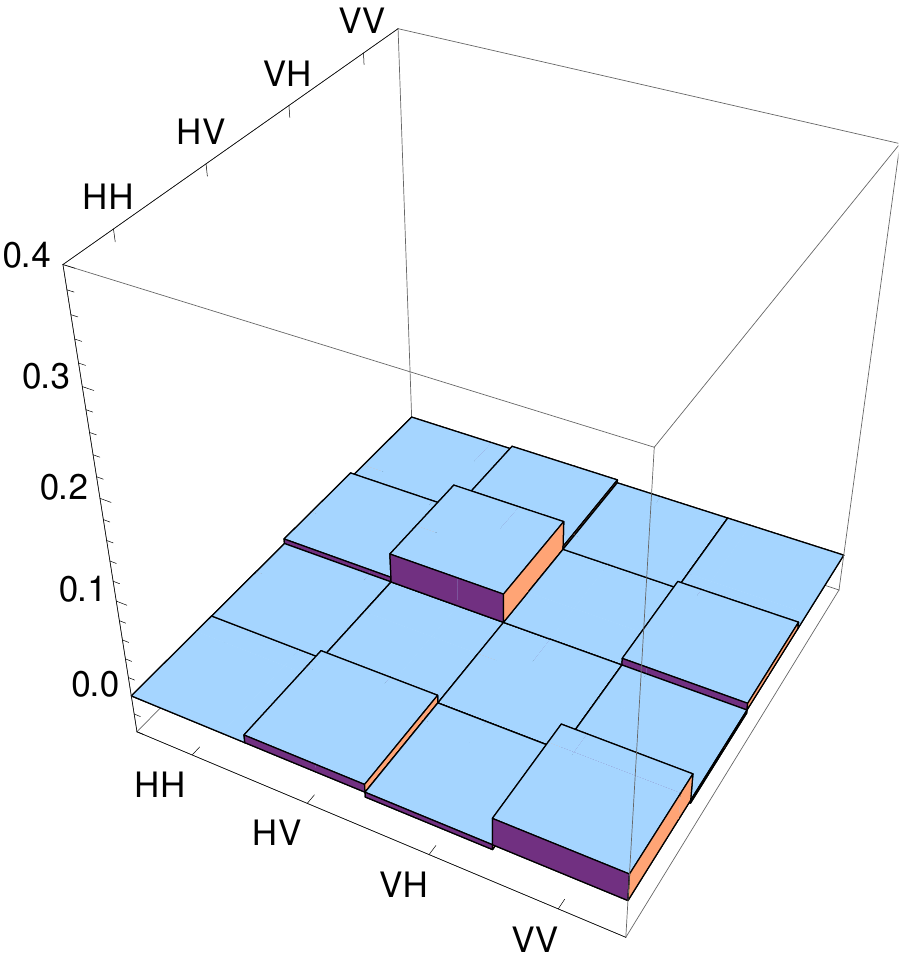}
\includegraphics[width=0.45\columnwidth]{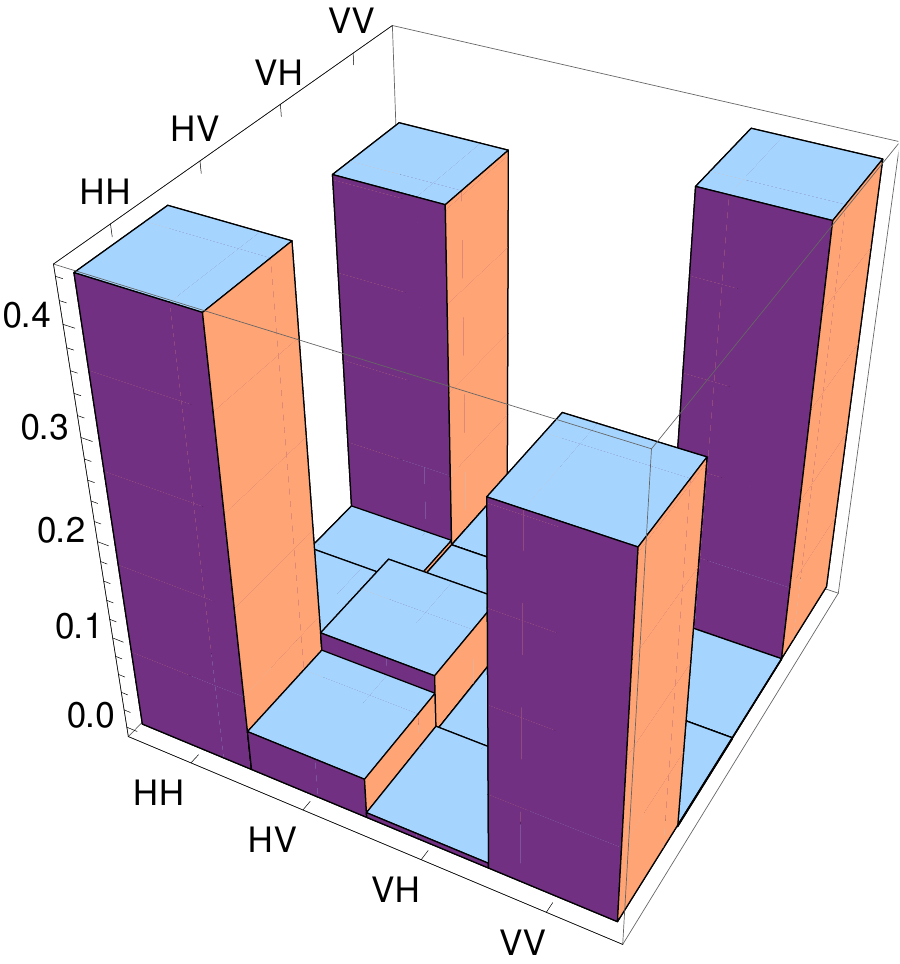}
\includegraphics[width=0.45\columnwidth]{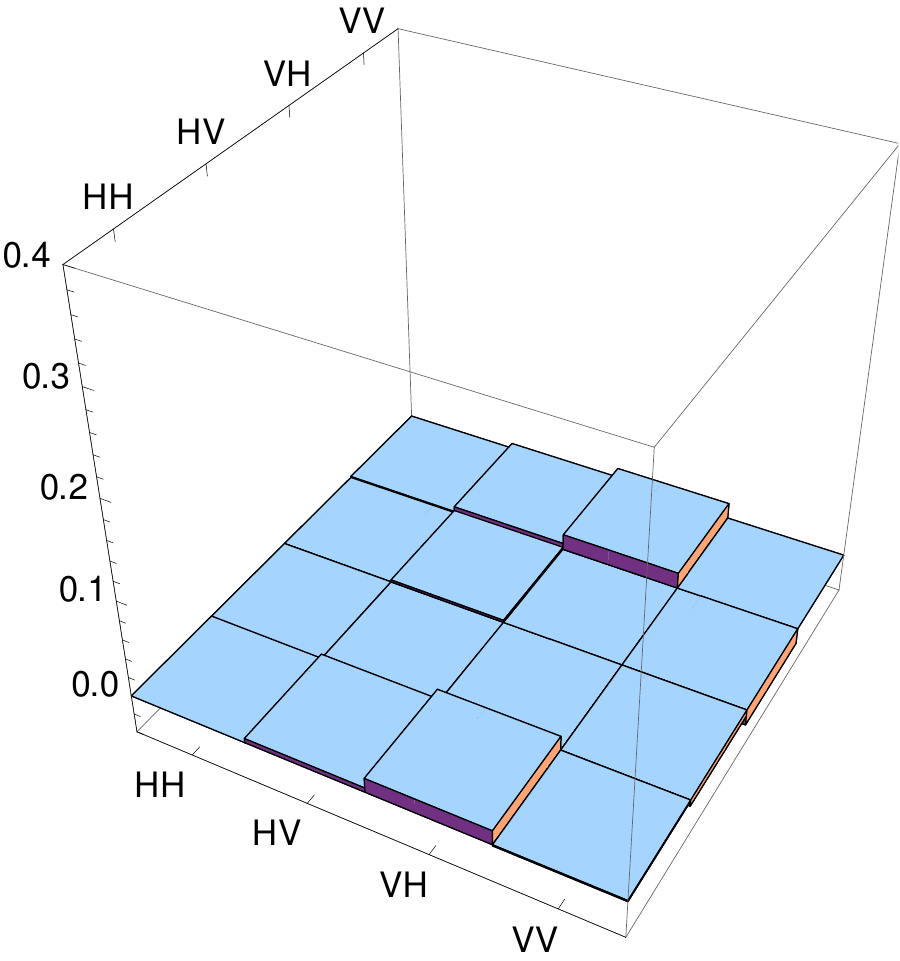}
\caption{Real (left) and imaginary (right) part of the reconstructed density matrices
of $|\psi^{+}\rangle$ (upper plot) and a $|\phi^{+}\rangle$ (lower plot) state}
\label{f2_tomo}
\end{figure}
\par
Two qubit quantum state tomography consists of set of projective 
measurements to different polarization states \cite{maxlik,Kwiat,cia09,OURs,yu2011}. 
This is achieved using quarter--wave and half--wave plates in both channels.  In
particular, a set of independent two-qubit projectors
$\ketbra{\Psi_{\nu}}{\Psi_{\nu}}$ with $\nu=0\dots 16$ were measured
\cite{Kwiat}.
The probabilities $p_{\nu}=\langle\Psi_\nu|\rho | \Psi_\nu\rangle$ are 
estimated by the number of coincidence counts $n_{\nu}$ obtained measuring 
the projector $P_{\nu}$. 
In order to enforce positivity of the reconstructed state, we employ a 
Maximum Likelihood estimation scheme, where the density
matrix is written as $\rho=T^{\dagger}T/\text{Tr}[T^{\dagger}T]$,
$T$ being a complex lower triangular matrix. We have 16 real variables
$t_j$ to be determined, with the physical density matrix given by
$\rho_L(t_1,t_2,\dots,t_{16})$. 
The likelihood function assesses how the reconstructed density matrix 
 $\rho_L(t_1,t_2,\dots,t_{16})$  reproduces the experimental data and 
 it is a function of both the data counts 
$n_{\nu}$ (proportional to $p_\nu$) and the coefficients 
$t_{\nu}$, $\mathcal{L}(t_1,t_2,\dots,t_{16};n_1,n_2,\dots,n_{16})$. 
In the Gaussian approximation \cite{Kwiat} the log-likelihood function
for a given data count set $\{n_{\nu}\}_{\nu=1}^{16}$ is given by
\begin{align}
 &\mathcal{L}(t_1,t_2,\dots,t_{16})=\nonumber\\
&N_T \sum_{\nu=1}^{16}\frac{(\bra{\Psi_{\nu}}\rho_L(t_1,t_2,\dots,t_{16})
\ket{\Psi_{\nu}}-n_{\nu})^2}{2\bra{\Psi_{\nu}}\rho_L(t_1,t_2,\dots,t_{16})\ket{\Psi_{\nu}}}
\end{align}
where $N_T=\sum_{\nu=1}^4 n_{\nu}$ is a constant proportional to the total number of runs.
By numerically maximizing the log-likelihood over the coefficients
$t_1,t_2,\dots,t_{16}$, we obtain the ML density matrix.
In Fig. \ref{f2_tomo} we report the the reconstructed density matrices
for the $|\psi^{+}\rangle$ (upper plot) and a $|\phi^{+}\rangle$ (lower plot)
states. We also performed tomography for Werner and phase-damped states
of the form (\ref{eq:MStateW}) and (\ref{eq:MStateDphi}) states with 
(theoretical) purity $\mu$ equal to $0.25$, $0.5$, $0.67$, $0.83$ and $1$. 
The resulting density matrices are very close to those expected
theoretically. In fact, the fidelities of the experimentally 
reconstructed two-qubit states to Werner and phase-damped models 
are very high: all being larger than 0.96, and most of them  
larger than 0.98. The same is true for the values of the purity, as 
obtained from the reconstructed density matrices.
Results are summarized in Tables \ref{t:expF} and \ref{t:expmu}.
Uncertainties are evaluated assuming that counts $n_{\nu}$ are Poissonian 
distributed, with mean equal to the experimental recorded value. 
Then, we numerically sample counts from Poisson distributions and
reconstruct the corresponding density matrices by maximum likelihood
method. In this way, we generate a sample of physical density matrices 
and for each one we compute the fidelity and the purity. The standard 
deviation associated to these values represents the uncertainty of these
quantities as estimated from tomographic reconstruction \cite{gum}.
\begin{table}[h!]
\begin{tabular}{|c|c|c|}
\hline
$\mu\left(\hbox{th}\right)$& 
$F \left(\rho_W^\phi\right)$ & $F\left(\rho_W^{\psi}\right)$\\
\hline
1 & 0.98 $\pm$ 0.02 &0.988 $\pm$ 0.004 \\
\hline
$0.83$ & 0.96 $\pm$ 0.01 &0.962 $\pm$ 0.006\\
\hline
0.66 & 0.96 $\pm$ 0.02 & 0.973 $\pm$ 0.02\\
\hline
0.50 & 0.99 $\pm$ 0.01 & 0.99 $\pm$ 0.02\\
\hline
0.25 & 0.991 $\pm$ 0.009 & 0.991 $\pm 0.009$\\
\hline
$\mu\left(\hbox{th}\right)$&
$F\left(\rho_D^{\phi}\right)$ & $F\left(\rho_D^{\psi}\right)$\\ 
\hline
1 & 0.98 $\pm$ 0.02 &0.988 $\pm$ 0.004 \\
\hline
0.83 &0.998 $\pm$ 0.002 & 0.995 $\pm$ 0.002 \\
\hline
0.66 & 0.997 $\pm$ 0.002 & 0.996 $\pm$ 0.002\\
\hline
0.50 & 0.997 $\pm$ 0.002 & 0.998 $\pm$ 0.002\\
\hline
\end{tabular}
\caption{Fidelity between the reconstructed states and the corresponding 
theoretical X-states belonging to the families of Werner and
phase-damped $\ket{\Phi^{+}}$ or $\ket{\Psi^{+}}$ states.}\label{t:expF}
\end{table}
\begin{table}[h!]
\begin{tabular}{|c|c|c|c|c|}
\hline
$\mu\left(\hbox{th}\right)$&
$\mu_T\left(\rho_W^{\phi}\right)$&
$\mu_X\left(\rho_W^{\phi}\right)$ & 
$\mu_T\left(\rho_W^{\psi}\right)$&
$\mu_X\left(\rho_W^{\psi}\right)$\\
\hline
1&0.96$\pm$0.04&0.996$\pm$0.002
&0.984$\pm$0.08&0.995$\pm$0.002\\
\hline
0.83&0.79$\pm$0.04&0.84$\pm$0.01
&0.82$\pm$0.03&0.87$\pm$0.01\\
\hline
0.66&0.66$\pm$0.04&0.68$\pm$0.02
&0.66$\pm$0.03&0.69$\pm$0.02\\
\hline
0.50&0.48$\pm$0.03&0.46$\pm$0.01
&0.47$\pm$0.02&0.52$\pm$0.01\\
\hline
0.25&0.259$\pm$0.006&0.25$\pm$0
&0.267$\pm$0.007&0.25$\pm$0\\
\hline
$\mu\left(\hbox{th}\right)$&
$\mu_T\left(\rho_D^{\phi}\right)$&
$\mu_X\left(\rho_D^{\phi}\right)$
&$\mu_T\left(\rho_D^{\psi}\right)$&
$\mu_X\left(\rho_D^{\psi}\right)$\\
\hline
1&0.96$\pm$0.04&0.996$\pm$0.002
&0.984$\pm$0.008&0.995$\pm$0.002\\
\hline
0.83&0.81$\pm$0.04&0.86$\pm$0.05
&0.79$\pm$0.02&0.82$\pm$0.01\\
\hline
0.66&0.63$\pm$0.02&0.67$\pm$0.03
&0.64$\pm$0.02&0.65$\pm$0.01\\
\hline
0.50&0.501$\pm$0.001&0.5$\pm$0
&0.504$\pm$0.003&0.502$\pm$0.002\\
\hline
\end{tabular}\caption{Purity of the reconstructed states as estimated
from tomography and from the X-state model.} \label{t:expmu}
\end{table}
\section{Estimation of discord}
\label{s:expD}
In order to estimate the value of discord, different techniques have
been employed and compared. The first method is to employ two-qubit 
tomography of the state to estimate its density matrix and then 
determine the quantum discord from its definition in Eq. \eqref{eq:Disc}. 
This is done in two steps: At first we evaluate the quantum mutual
information in Eq. (\ref{eq:QMI}), which requires the reconstructed 
density matrix only. Then the classical correlations are computed using 
Eq. \eqref{eq:CCor}. The minimization over all possible measurements
is performed numerically, without assuming any specific form for the
state. In fact, despite the high fidelity, the experimentally reconstructed 
density matrices do not have an exact X-shape, as shown in Fig.
\ref{f2_tomo}. This is due to fluctuations in the coincidence counts.
In order to evaluate how this fluctuations propagate into fluctuations
of the discord we have again assumed that counts $n_{\nu}$ are Poissonian 
distributed and generated a set of physical density matrices by 
Monte Carlo sampling of the coincidence counts.
The standard deviation associated to these
discord values is the uncertainty associated with the estimated value of
discord via tomographic technique. 
\par
\begin{figure}[h!]
\includegraphics[width=0.45\columnwidth]{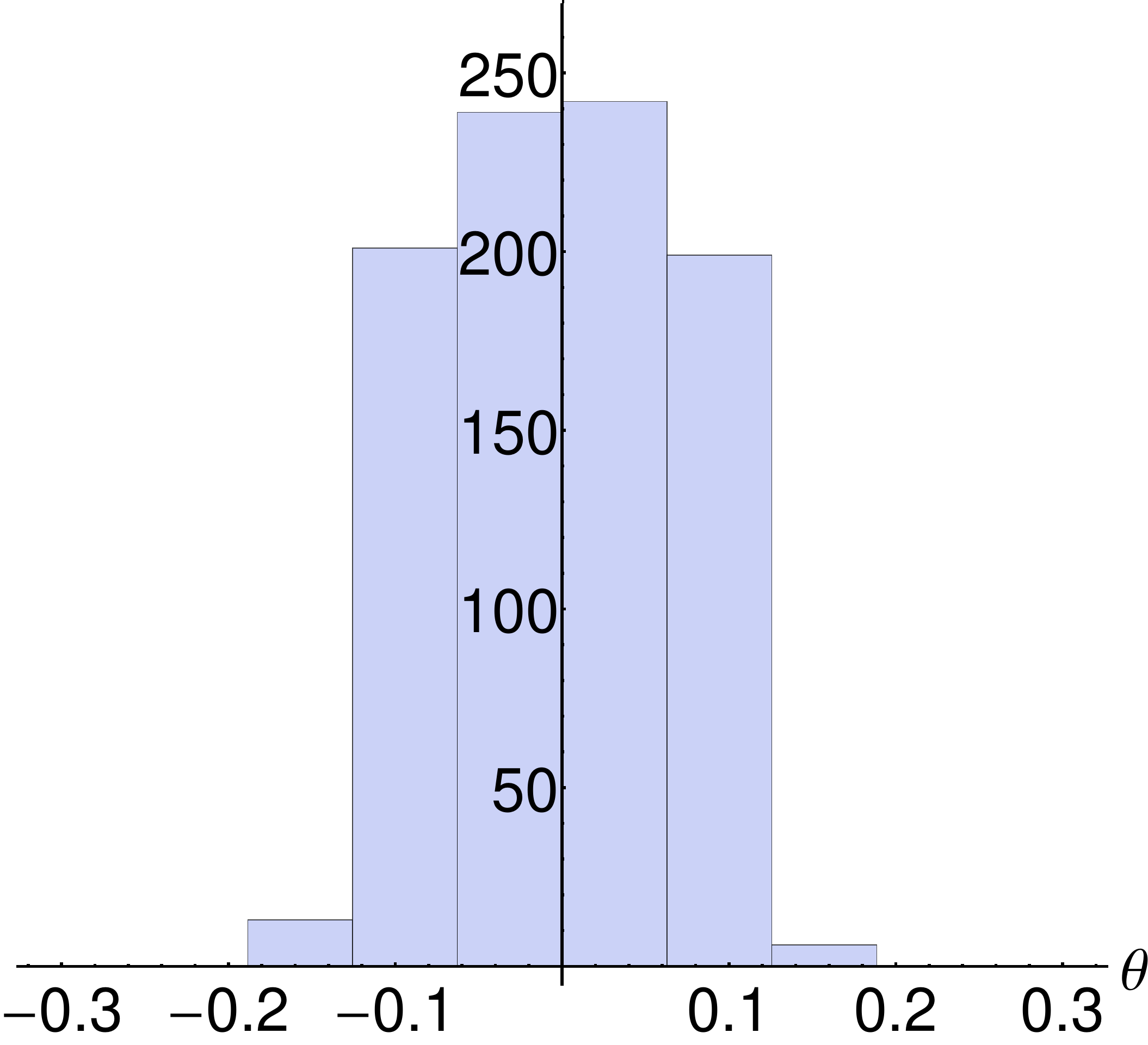}
\includegraphics[width=0.45\columnwidth]{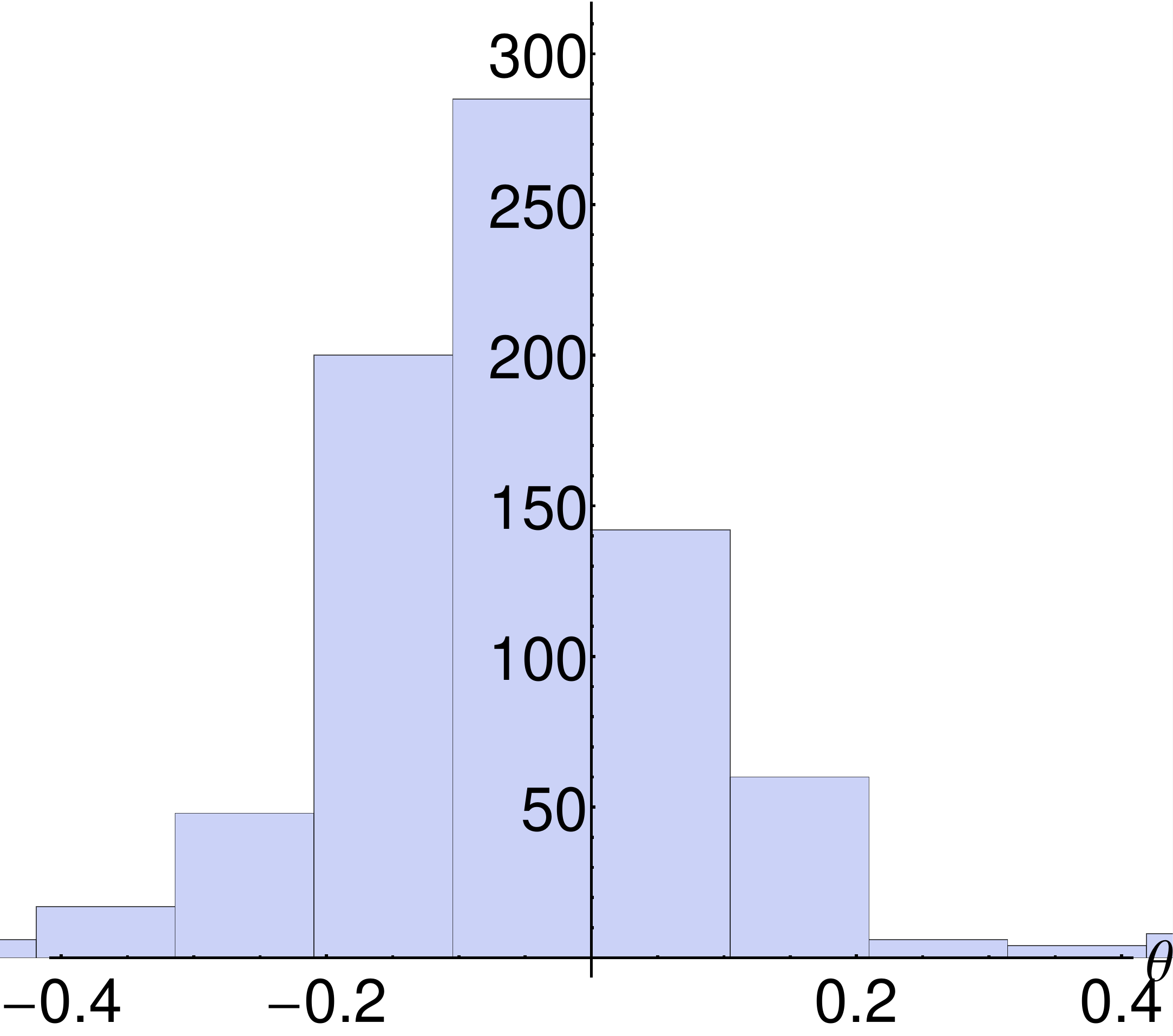}
\caption{Optimal measurement to achieve classical correlations for some
of the states generated in our experiments. We report the distribution
of optimal angles $\theta$ for a set of 900 density matrices 
obtained by Monte Carlo sampling of coincidence counts. The left panel
is for a phase-damped $|\Psi^+\rangle$ state with purity $\mu=0.66$ and 
the right panel for an unperturbed $|\Phi^+\rangle$ state.}
\label{f3_optm}
\end{figure}

As an alternative method to evaluate discord, we may 
use partial tomography as follows. The evaluation of the quantum 
mutual information (\ref{eq:QMI}) is done as above. Then, we estimate 
the classical correlations (\ref{eq:CCor}) explicitly performing the optimal 
measurement and using partial tomography to determine the conditional
states and in turn their entropy. To this aim, the quarter-wave and half-wave plates 
(QWP$_1$, WP$_1$) were used in one transmitted arm after the beam splitter 
to perform tomographic procedures only on the single photon state.  
No transformation  was performed in reflected arm, that is, QWP$_2$ and WP$_2$ 
were removed from setup. As mentioned in the introductory Section, 
the optimal projection operator, minimizing the sum in Eq.~(\ref{eq:CCor}), 
is $\sigma_z$ for both the families of mixed states \cite{Luo}.
Of course this technique is less general than the previous one since,
strictly speaking, $\sigma_z$ is the optimal measurement only for states
that are exactly Werner or phase-damped states. On the other hand, the
large values of fidelity suggest that optimal measurement are not very
far from the theoretical one, as well as the corresponding values of the
classical correlations. We have validated this assumption {\em a
priori}, by numerical evaluation of the optimal measurement (see the
following paragraph and Fig.
\ref{f3_optm}), and {\em a posteriori}, by comparing the resulting
values of the discord with the results obtained from full tomography 
of the two qubit states. 
\par
The results about the optimization of
measurement are summarized in Fig. \ref{f3_optm} for 
some of the states of Table \ref{t:expF}. We show the 
optimal measurement angle $\theta$ 
for the set of physical density matrices obtained 
by Monte Carlo sampling of the coincidence counts. 
The optimization has been performed over projective measurements
of the form $\sigma(\theta,\phi) = \Pi_0-\Pi_1$, where 
$\Pi_k = \frac12
({\mathbb I} + \mathbf{n}_k.\mathbf{\sigma})$, $\mathbf{n}_0 =
(\sin\theta\cos\phi, \sin\theta\sin\phi,\cos\theta)$ and
$\mathbf{n}_0\perp\mathbf{n}_1$. As it is apparent from the plot
the optimal measurement is close to $\sigma_z$ (i.e. $\theta=0$)
for the entire sample, and the fluctuations are small.
\par
The excellent agreement of the reconstructed states with the 
theoretical (Werner and phase-damped) models also suggests a third
method to access discord: Upon assuming that our generated states 
belong to the families of single-parameter mixed  states $\rho_W(p)$ 
and $\rho_D(p)$, we may estimate the value of discord using Eqs. 
(\ref{dw}) and (\ref{dd}), i.e. by estimating only the single parameter 
$p$. An estimator for this quantity may be determined using 
any function linking the mixing $p$ to the number 
of coincidence counts. Actually, we dispose of four of these functions 
for Werner states, and of two for decohered states. 
They are given by
\begin{align}
&\frac{1+(-1)^{\nu}p}{4}=\frac{n_{\nu}}{N_T}\quad
\nu=1,2,3,4\quad \text{for }\rho_{W}(p) \hfill ,\\
&\frac{1-(-1)^{\frac{\nu}{2}}p}{4}=\frac{n_{\nu}}{N_T}\quad
\nu=10,16 \quad\text{for }\rho_{D}(p)\hfill .
\end{align}
We have used the above relations to build a randomized estimator for
$p$ for both families and have evaluated their precision using Monte
Carlo sampling of data (again assuming a Poissonian distribution 
for counts).
\section{Results and discussion}
\label{s:res}
In Table \ref{t:disc} we report the estimated values of 
quantum discord for depolarized and phase-damped $|\Phi^+\rangle$ 
and $|\Psi^+\rangle$ states, together with experimental 
uncertainties. We compare the values obtained via total and partial
tomography with the determination achieved by assuming the
single-parameter X-state model. 
\begin{table}[h!]
\begin{tabular}{|c|c|c|c|c|}
\hline
$\left[\rho_W^{\phi}\right]$ &$\mu\left(\hbox{th}\right)$
&TT & PT&$\rho_W^{\phi} (p)$ \\
\hline
&1&0.9$\pm$0.1 &0.9$\pm$0.1 &0.997$\pm$0.002\\
\hline
&0.83&0.59$\pm$0.07&0.7$\pm$0.1&0.76$\pm$0.02\\
\hline
&0.66&0.47$\pm$0.05&0.6$\pm$0.1&0.56$\pm$0.02\\
\hline
&0.50&0.24$\pm$0.03&0.35$\pm$0.09&0.29$\pm$0.02\\
\hline
&0.25&0.009$\pm$0.007&0.04$\pm$0.04&0$\pm$9$\;10^{-7}$\\
\hline
\hline
$\left[\rho_D^{\phi}\right]$ &$\mu\left(\hbox{th}\right)$ &TT & PT&$\rho_D^{\phi}(p)$\\
\hline
&0.83&0.53$\pm$0.08&0.53$\pm$0.08&0.6$\pm$0.1\\
\hline
&0.66&0.20$\pm$0.04&0.20$\pm$0.04&0.26$\pm$0.05\\
\hline
&0.50&0.013$\pm$0.006&0.01$\pm$0.01&0$\pm$1$\;10^{-8}$\\
\hline
\hline
$\left[\rho_W^{\psi}\right]$ &$\mu\left(\hbox{th}\right)$ &TT & PT&$\rho_W^{\psi}(p)$\\
\hline
&1&0.94$\pm$0.03&0.94$\pm$0.03&0.989$\pm$0.005\\
\hline
&0.83&0.67$\pm$0.05&0.70$\pm$0.09&0.79$\pm$0.02\\
\hline
&0.66&0.47$\pm$0.04&0.5$\pm$0.1&0.57$\pm$0.02\\
\hline
&0.50&0.25$\pm$0.04&0.30$\pm$0.08&0.37$\pm$0.02\\
\hline
&0.25&0.009$\pm$0.007&0.04$\pm$0.02&0$\pm 1\;10^{-10}$\\
\hline
\hline
$\left[\rho_D^{\psi}\right]$ &$\mu\left(\hbox{th}\right)$ &TT & PT&$\rho_D^{\psi}(p)$\\
\hline
&0.83&0.49$\pm$0.04&0.48$\pm$0.04&0.53$\pm$0.03\\
\hline
&0.66&0.21$\pm$0.03&0.20$\pm$0.03&0.23$\pm$0.02\\
\hline
&0.50&0.008$\pm$0.005&0.016$\pm$0.001&0.003$\pm$0.003\\
\hline
\end{tabular}\caption{Estimation of discord the families of 
$\phi$- and $\psi$-states generated in our experiments. We report the
determination obtained from total (TT) and partial (PT) 
tomography, and from the single-parameter X-state model.}\label{t:disc}
\end{table}
\par
\begin{figure}[h!]
\includegraphics[width=0.49\columnwidth]{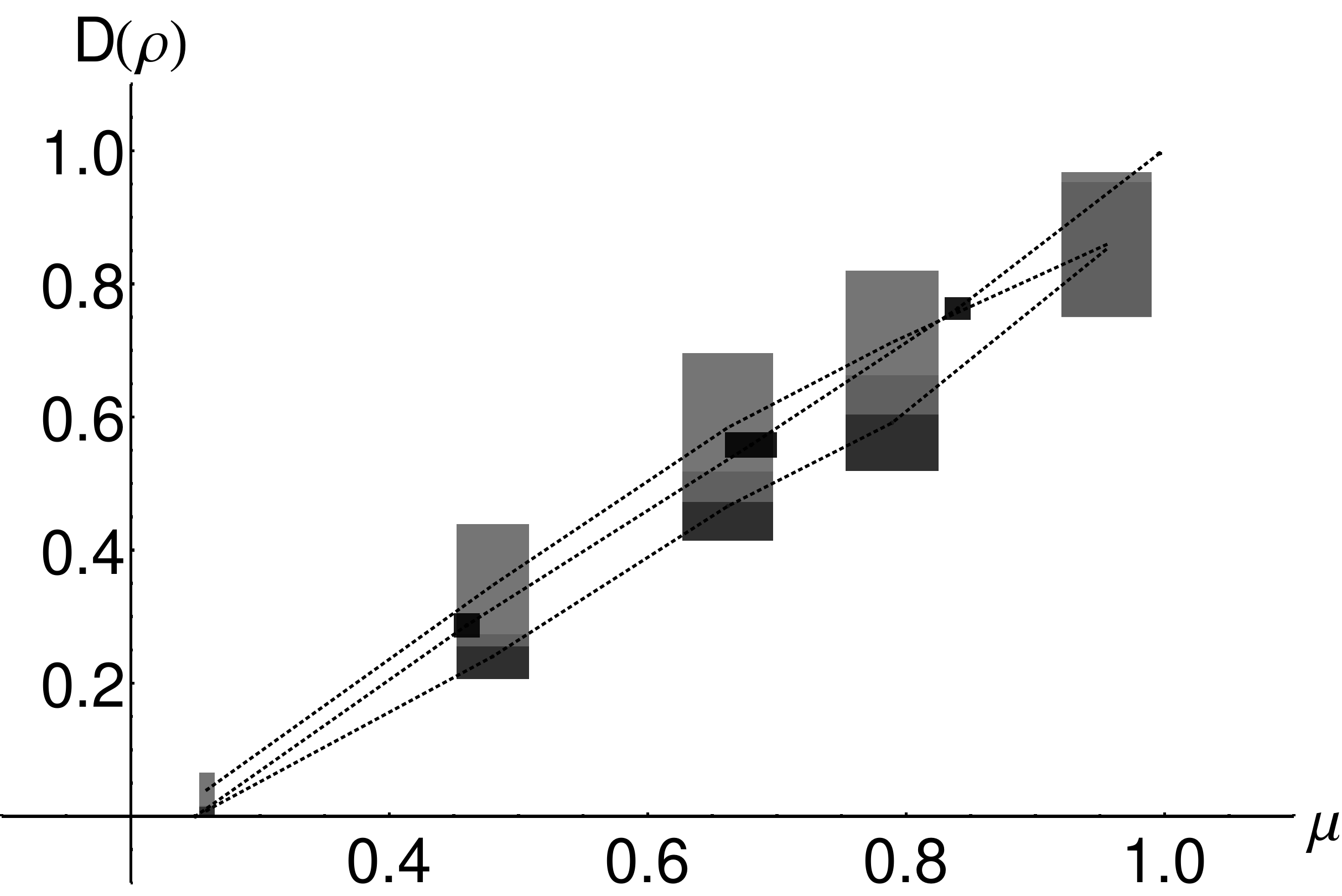}
\includegraphics[width=0.49\columnwidth]{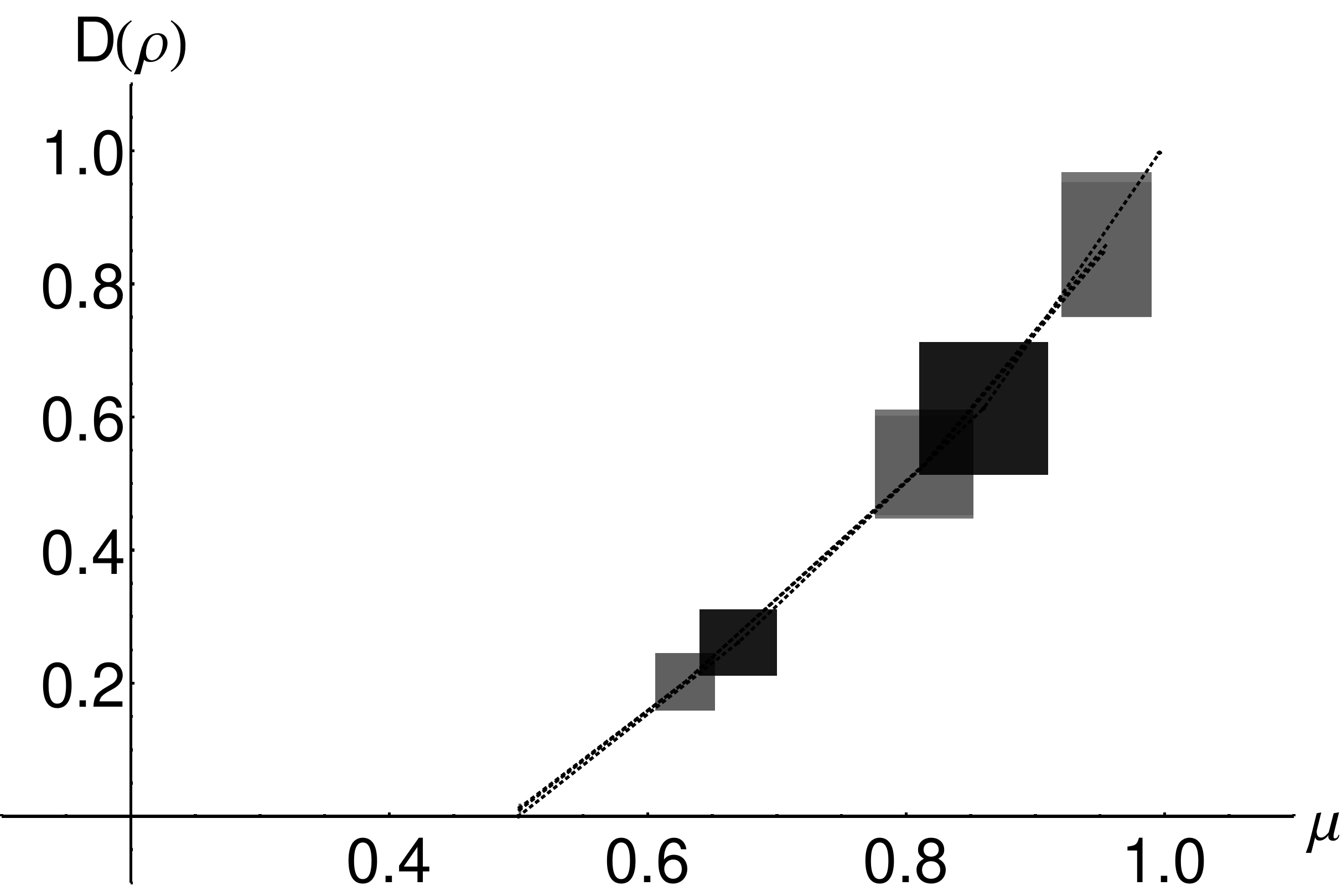}
\includegraphics[width=0.49\columnwidth]{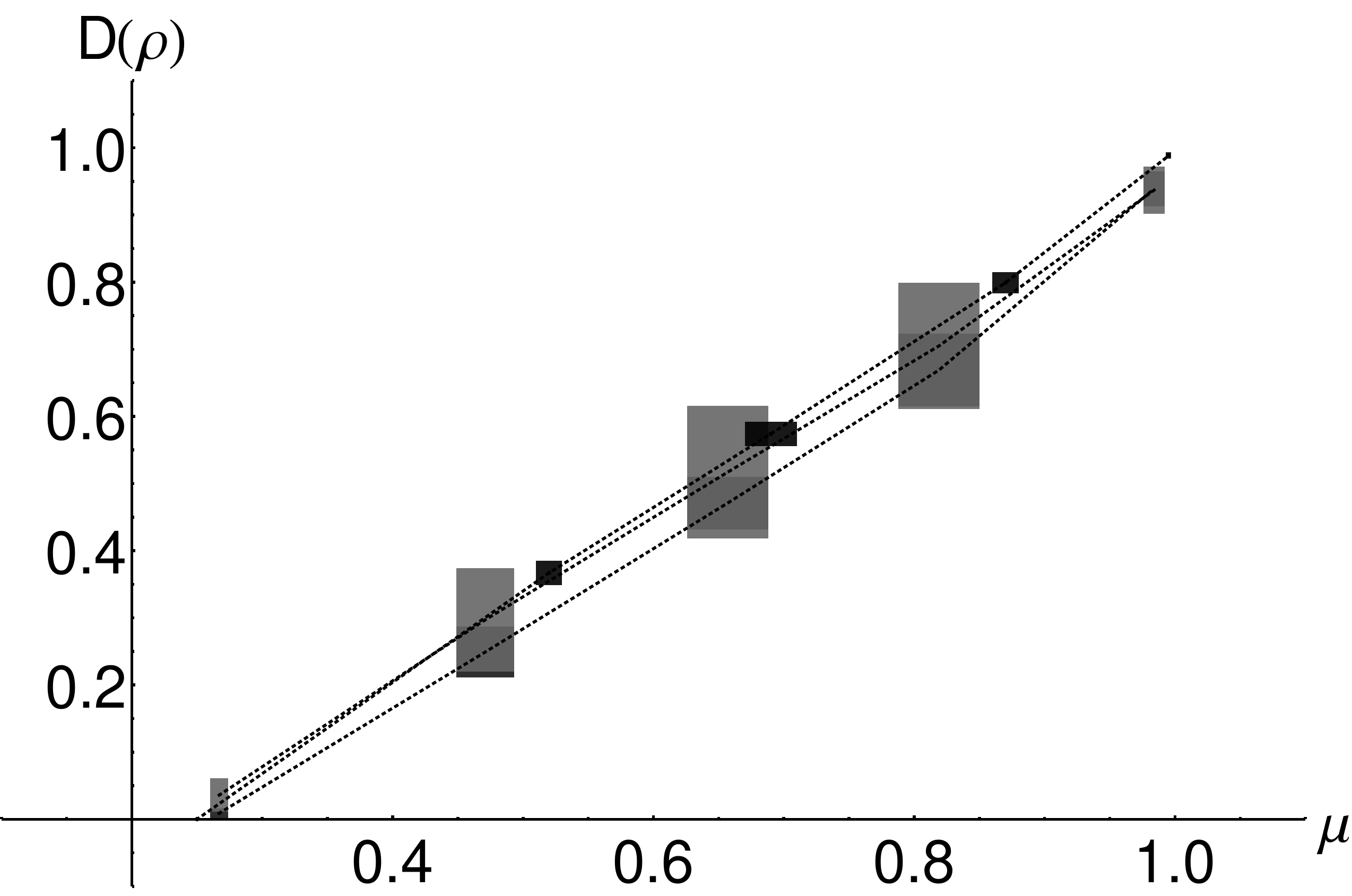}
\includegraphics[width=0.49\columnwidth]{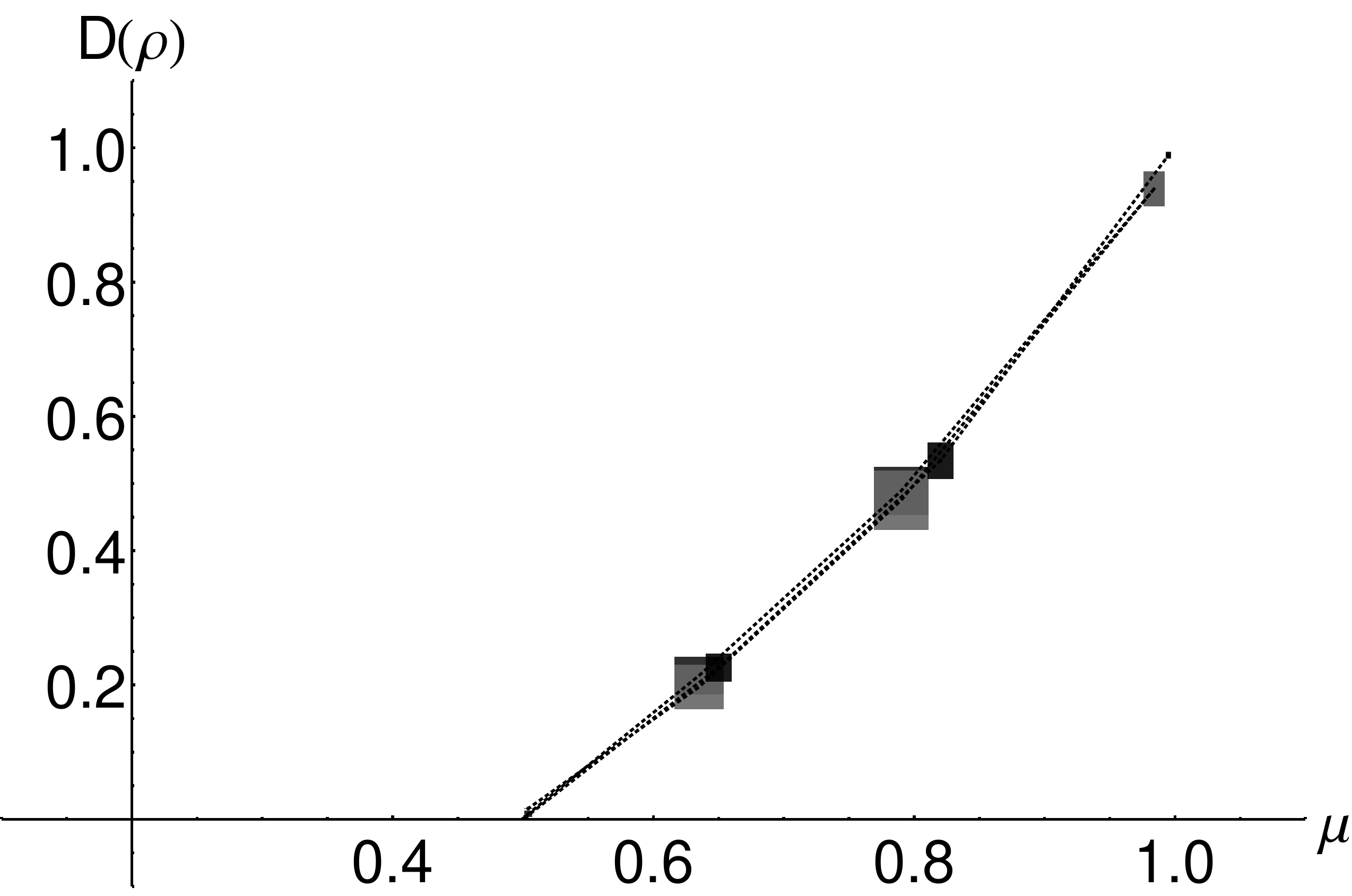}
\caption{(Upper panels): Estimation of discord from total (dark gray)
and partial (light gray) tomography and for the X-state model (black),
with experimental errors as a function of the purity of the assumed
Werner(left) and phase-damped (right) states obtained from state
$\ket{\Phi^{+}}$.  (Lower panels): the same for $\ket{\Psi^{+}}$
\label{f4_disc}} \end{figure}

In order to better compare these values, we also summarize them 
in Fig. \ref{f4_disc}. As it is apparent from the plots the two 
tomographic determinations are in good agreement within the 
uncertainties. 
More specifically, partial tomography (light
gray squares) slightly overestimates the discord compared to total
tomography (dark gray squares) for Werner states, while there is 
no appreciable difference for phase-damped states. 
On the other hand, despite the high-fidelity of the reconstructed states
to the single-parameter (Werner or phase-damped) models, the discord
estimated within this assumption (black square) is not always compatible
with the corresponding tomographic determination.  In particular,
results from full tomography and the single-parameter models are never
statistically compatible, and this happens also for partial tomography
in some cases. This discrepancy questions the usefulness of fidelity in
assessing quantum correlations. In fact, even if the reconstructed
states have a very high fidelity to theoretical states $\rho_W(p)$ and
$\rho_D(p)$, the estimated values of discord may be very different. The
motivation behind this behavior is twofold.  On the one hand, our
states are not genuine X-states, despite the high value of fidelity. On
the other hand, states that are very close in terms of fidelity may have
very different values of discord.  This argument, together with the fact
that an analytic expression for the discord of arbitrary two qubits
states cannot be obtained \cite{gir11}, leads to the conclusion that the
only way to estimate the discord is through a tomography process, which
itself is, in general, an intrinsically noisy procedure \cite{tn,adv}.
\par
\begin{figure}[h!]
\includegraphics[width=0.9\columnwidth]{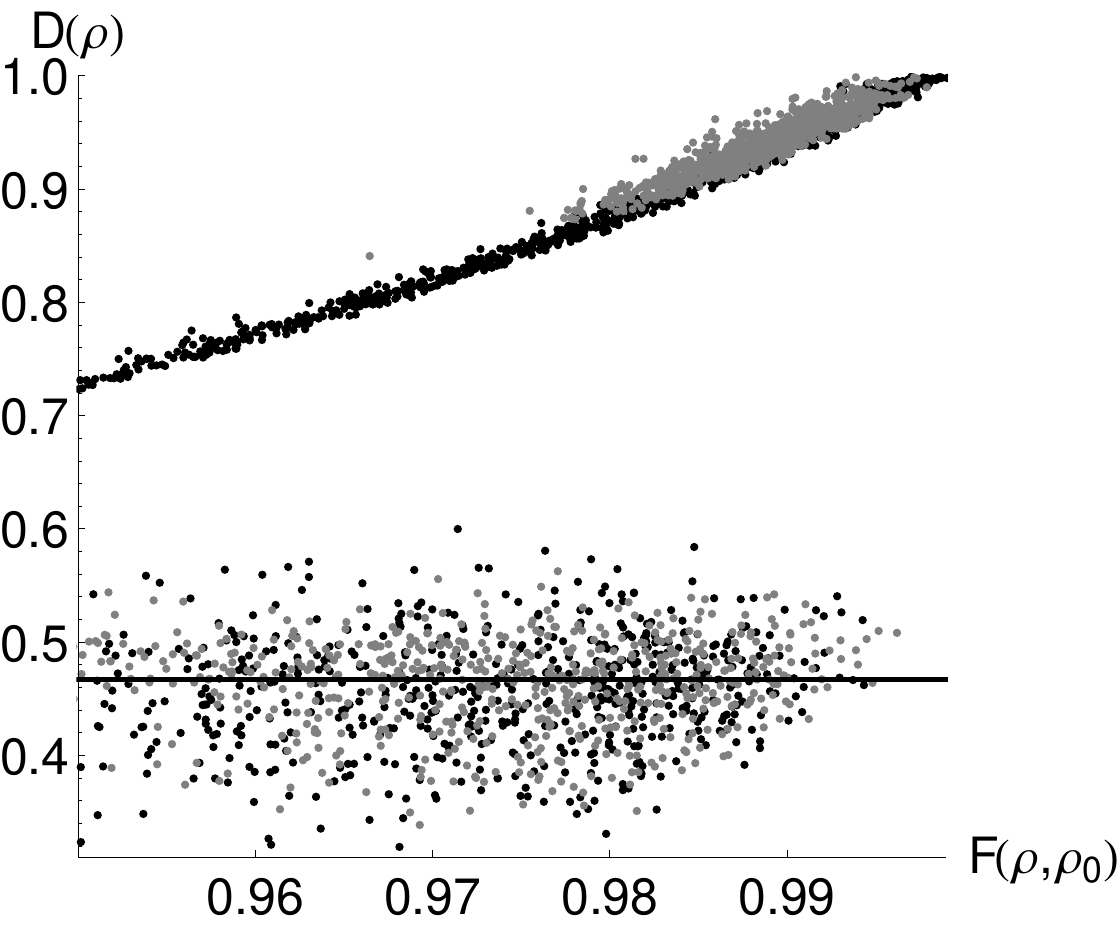}\quad
\caption{Distribution, in the $F$-$D$ (fidelity-discord) plane, 
of random states close, in terms of fidelity, to some states of Table \ref{t:expF}.
In the upper area of the plot we have gray points for states close
to $|\Psi^+\rangle$ and black points for states close to
$|\Phi^+\rangle$. In the lower area we have the same for states close
to states $\varrho_W^\psi$ and $\varrho_W^\phi$ (purity $\mu=0.66$)
respectively.}\label{fig5}
\end{figure}

In order to illustrate explicitly the behavior of quantum discord
for neighbouring states we have generated a set of random states 
(by Monte Carlo sampling starting from the recorded data set) 
in the vicinity of the reconstructed states, which are used as 
fixed reference states. We then compute both the fidelity
$F(\varrho,\varrho_0)$ between each random state and the reference one, 
as well as the discord of the generated state. Few examples of the resulting
distributions are shown in Fig. \ref{fig5}. As it is apparent from the
plot, states close, in terms of fidelity, to Werner states 
$\varrho_W^\phi$ or $\varrho_W^\psi$ with purity $\mu=0.66$,  
are also close in terms of discord, whereas there 
is a large fraction of states with a  very high fidelity to
$|\Psi^+\rangle$ or $|\Phi^+\rangle$ having a very different value of 
discord. In other words, discord is 
a highly non-linear function of the fidelity parameter, such that small
deviations from its value may lead to very different values of 
discord.
\section{Conclusions}
\label{s:out}
In this paper we have  addressed the experimental estimation of quantum
discord for systems made of two correlated polarization qubits.  We have
compared the results obtained using full and partial tomographic methods
and have shown that they are in good agreement within the experimental
uncertainties. We have also computed the fidelity of the reconstructed
states to suitable Werner and phase-damped states, and found very high
values. This fact would in principle allows one to estimate quantum
discord by assuming a single-parameter X-shape form for the
reconstructed states, and extracting the mixing parameter from the data.
Nonetheless, using the analytic expression for discord of Werner and
decohered states, we found results that are statistically not compatible
with the tomographic ones. This means that the assumed model is not
usable, and that estimation of entropic discord for polarization qubits
necessarily requires a tomographic  reconstruction. 
\par 
Indeed, states that are very close in terms of
fidelity may have very different values of discord, i.e.  discord is a
quantity very sensitive to small perturbations.  In fact, our states are
not genuine X-states, despite the high value of fidelity, and this fact,
together with the lack of an analytic expression for the quantum discord
of a generic two-qubit state, leaves tomographic reconstruction as the
sole reliable method to estimate quantum discord of polarization qubits.
Our results also question the relevance and the role of fidelity as a
tool in the evaluation of quantum correlations.
\section{Acknowledgments}
This work has been supported by MIUR (FIRB LiCHIS-RBFR10YQ3H) 
and by Compagnia di San Paolo. 
 
\end{document}